\documentclass[useAMS,usenatbib,usegraphicx]{mn2e}

\def\la{\raise.5ex\hbox{$<$}\kern-.8em\lower 1mm\hbox{$\sim$}}
\def\ma{\raise.5ex\hbox{$>$}\kern-.8em\lower 1mm\hbox{$\sim$}}

\def\msol{M$_{\odot}$ }

\def\kms{$\rm km\, s^{-1}$}
\def\cm3{$\rm cm^{-3}$}

\def\Vs{$\rm V_{s}$~}
\def\n0{$\rm n_{0}$}
\def\B0{$\rm B_{0}$}

\def\Tgr{$\rm T_{gr}$}

\def\erg{$\rm erg\, cm^{-2}\, s^{-1}$}
\def\mum{$\mu$m~}

\def\L12{L$_{12\mu m}$~}
\def\F12{F$_{12\mu m}$~}
\def\agr{a$_{gr}$}
\def\Hb{H$\beta$~}

\def\Tef{T$_{eff}$~}
\def\roi{R$_{[OI]}$~}

\title[The LLAGN in the GC]{The  LLAGN in the centre of the  Galaxy}
\author[M. Contini]{M. Contini$^$}
  
\author[M. Contini]{ M. Contini$^{ }$ 
\\
$^{ }$School of Physics and Astronomy, Tel Aviv University, Tel Aviv
69978, Israel \\
}

\begin{document}

\date{Accepted: Received ; in original form 2010 month day}

\pagerange{\pageref{firstpage}--\pageref{lastpage}} \pubyear{2009}

\maketitle

\label{firstpage}

\begin{abstract}
The  observations of FIR line and continuum spectra throughout the Galactic centre and in some
regions of the disc  are  analysed in order to   determine the physical conditions (densities, shock velocities,
 radiation parameters, etc.) and the relative abundances of some elements (C, N, and O).  
The models account for the coupled effect of photoionization and shocks.
Consistent model calculations  of the line and continuum spectra show that, 
although the radiation from the  stars dominates,
 an active galactic nucleus (AGN) is clearly present, with    a radiation maximum  in the Sgr A* region.
The flux,  similar to that found  for  low luminosity AGN (LLAGN), is  lower by  a factor of $\sim$ 100  
than that  of AGN.  
Gas densities in the downstream line emission region range between 100 and 2000 \cm3, the shock velocities
between 50 and 300 \kms.  
Densities of $\sim$ 5 10$^6$ \cm3, close to the Sgr A* black hole, 
lead to   self-absorption of  free-free radiation in the radio frequency range,
while X-ray data are explained by shock velocities of $\sim$ 3000 \kms.
A magnetic field  of  $\sim$ 10 $^{-4}$ gauss shows relatively small fluctuations throughout the Galactic centre.
The dust-to-gas ratios range between 3 10$^{-15}$ and $\leq$ 10$^{-13}$ by number. Lower values are found
far from the centre, suggesting
that  N and O, which are depleted from the gaseous phase, are included into molecules rather than
trapped into grains. 

\end{abstract}


\section{Introduction}

Acceleration measurements of stellar orbits
near the radio source Sgr A*  suggest the existence of a supermassive black
hole  at the Galactic centre (Ghez et al. 2003; Sch\"{o}del et al. 2003).
Recently, Genzel et al (2010) showed from the analysis of more than two dozen orbits
and from measurement of the central compact radio source Sgr A*, that this  must be 
a massive black hole (BH) of $\sim$ 4.4 10$^6$ \msol.

In particular, the   BH in the  central Sgr A West HII region  (Ghez et al.
 2005; Eisenhauer et al. 2005) is coincident with the radio source Sgr A*
and with   a cluster of massive stars.
Two  other  clusters of  young massive stars and massive molecular clouds
(Sch\"{o}del et al. 2006) appear  near the Galactic Centre  (GC),  the Arches Cluster and the Quintuplet Cluster
located $\sim$  25 pc away in the plane of the sky.
The very massive Arches Cluster (Nagata et al 1995 and Cotera et al. 1996)
 of young stars  heats and ionizes the  region of the
Arched Filaments. The Quintuplet Cluster ionizes the Sickle
and  affects the clouds in  extended regions  including  the Bubble.

The  Galactic central regions are  strongly obscured  to UV-optical spectroscopic observations. 
The  morphology of the regions near the GC,   recently  revealed  
at   radio frequencies,  shows
a  turbulent regime  on a large scale  (Simpson et al. 2007, fig.1).

The GC contains within  only 5 arcmin (12 pc) a massive BH 
surrounded by a rotating circumnuclear disc of dust and gas, HII regions, 
massive stellar clusters, two SNRs, and two giant molecular clouds (Amo-Baladr\'{o}n et al 2011).

The BH is  surrounded by gas and dust. The gas is 
photoionized by a dense stellar population  from the central cluster. Other stars
are distributed  throughout the central region  and  the disc.
Therefore the GC shows a composite nature of   AGN  and   starburst galaxy.
Moreover,  turbulent dynamical motions (e.g. arc-shaped gas streamers with a minispiral morphology)
imply a shock dominated  regime.

     The GC has been used as a test  for many models
that attempt to explain the physics of low luminosity AGN (LLAGN) in the last years
(e.g. Anderson et al. 2004). 
 In fact, the luminosity of Sgr A* is extremely low, even compared with other LLAGNs.

In AGNs as well as in  low-luminosity objects, such as LLAGNs  and LINERs
(low-ionization nuclear emission regions)  starbursts coexist with an active nucleus, affecting
 the spectra  with comparable importance (Contini et al 2002, Contini 1997, 2004a).
Modelling the X-ray-IR continuum correlation in a large sample of AGN,
Contini et al. (2003)  confirmed that LLAGNs  appear in the low-luminosity tail of AGN.
This is explained by the  relatively low flux from the active nucleus, roughly by two orders of magnitude
lower than found in the NLR of Seyfert galaxies (Contini 2004a). Moreover the
velocity field is  relatively low.

 Spitzer observations 
 of   IR spectra  (Simpson et al 2007 and references therein)  near the GC were analyzed   by
 Contini (2009, hereafter Paper I), leading to a detailed  picture of  gas and dust structures  near the GC.
The  modelling revealed the results of  ionization and heating   of the gaseous clouds by the 
young massive stars in  the Quintuplet Cluster, the Arches Cluster, etc. 
The presence of shocks generated by the  turbulent regime observed in these regions, definitively explained
 the NIR spectra. The comparison of  line ratios calculated by the models with those observed  in the IR 
was constrained by  a sufficiently high  number of ions.
The models accounted for the coupled effect of photoionization by a black body (BB) flux from the stars and shocks.

The models, which   explained the observed  IR spectra, were used to calculate
the UV and optical  lines which cannot be observed because strongly obscured 
(Contini \& Goldman 2010, hereafter Paper II). 
Comparison with observations  of active galaxies
(starbursts, AGN, LINERs, LLAGNs, etc) confirmed that the spectra near the GC are similar to those
dominated by a BB flux. However, an active nucleus  was not excluded.
Dust details emerged from the consistent modelling of the spectral energy distribution (SED) of the continuum
in the different regions.

The  results of Papers I and II  show that  some line ratios from regions near the GC are similar to 
those of  low luminosity
active galaxies, but the  gaseous clouds  are strongly fragmented by the underlying turbulence (Paper II).

\begin{figure*}
\begin{center}
\includegraphics[width=0.72\textwidth]{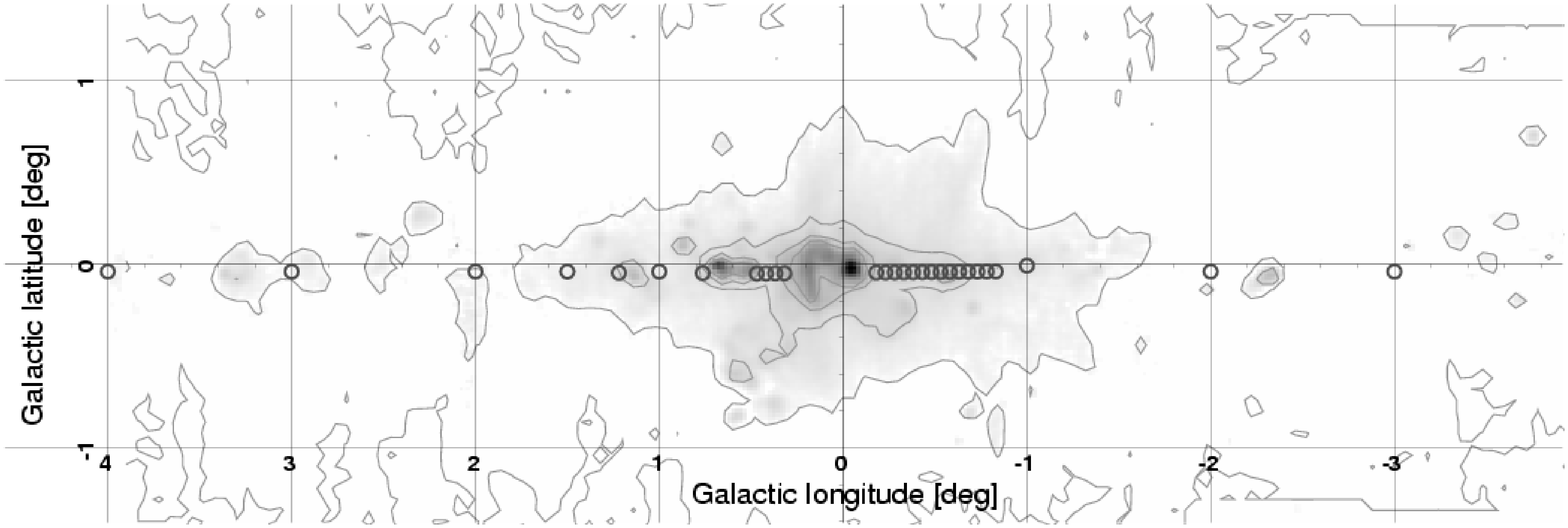}
\includegraphics[width=0.88\textwidth]{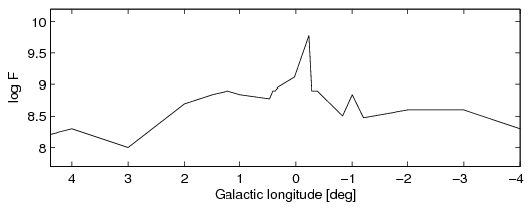}
\caption{top diagram : ISO/LWS observational points in the Galactic centre region,
plotted on the 10 GHz radio continuum map (Handa et al 1987),
adapted from Yasuda et al (2008, fig. 1).
Bottom diagram : the  calculated radiation flux $F$ corresponding
to power-law dominated models.
}
\end{center}
\end{figure*}

In the very central region of the Galaxy, there is
 observational evidence for  star-formation activity. 
The star-forming clouds, such as the Sgr B2, Sgr C, Sgr D, and the 50 \kms clouds in the Sgr A, 
show that  the star-formation activity is  high (Yasuda et al 2008).

Far-IR (FIR) spectra were observed in different regions of the GC.
Yasuda et al (2008)  report  on spectroscopic observations 
with the Long-Wavelength Spectrometer (LWS) on board the Infrared Space Observatory (ISO). 
Owing to the wide wavelength coverage (43 to 197 $\mu $m) of the ISO/LWS,  they could study multiple 
fine-structure lines at the same time, which  were used to determine the physical conditions 
of the interstellar medium (ISM). 
Previous ISO observations in the GC  were reported by many authors, 
 including  H II regions (Rodr\'{i}guez-Fern$\acute {\rm a}$ndez et al. 2004), 
giant molecular clouds (Lis et al. 2001) and discrete objects near the GC 
(Cotera et al. 2005; Goicoechea et al. 2004). 

Most of these observations focus on local regions, whereas  Yasuda et al. observations cover much larger 
GC regions. The targets of   Yasuda et al   are not particular regions, 
such as active H II regions, but   typical regions in the GC least 
contaminated by strong discrete sources.
Rodr\'{i}guez-Fern$\acute{\rm a}$ndez et al. (2004)  observed  similar regions. 

Yasuda et al  attempted to characterise the difference of general clouds in the GC 
region from those in the disc, and therefore  selected  GC
 as well as disc positions.  Rodr\'{i}guez-Fern$\acute {\rm a}$ndez et al. concentrated on the GC region.
Unfortunately, the   lines observed by  Rodr\'{i}guez-Fern$\acute {\rm a}$ndez et al.  do not 
constrain the models. 

Goicoechea et al. (2004) too observed the Sgr B2 region in the FIR fine structure emission lines, 
[N II], [N III], [O III], [C II], and [O I] and concluded that the 
local radiation field in the Sgr B2 region is characterized by a hard ionizing continuum typical of an O7 
star with a radiation effective temperature  $T_{\rm eff}$ of  $\sim$ 36 000 K.
Many  of the line intensity fluxes  are upper limits, therefore  the
spectra  cannot by used  for a detailed modelling (see, however, the  next sections).

In this paper we investigate the  GC regions and the  few disc ones  by modelling the 
 ISO/LWS archived data presented by Yasuda et al. (2008),
 accounting for both the star and the AGN characteristics of the GC.
First we will calculate the physical conditions of the emitting gas   
 adopting a BB dominated radiation flux
corresponding to  $T_{\rm eff}$  $\sim$ 36 000 K.
Then, we will  examine the AGN calculating the spectra by a power-law (pl) radiation flux
with  suitable spectral indices.

 According to Lis et al (2001) the harsh environment in the GC region (central $\sim$ 200 pc of the Milky Way),
the region referred to as the central molecular zone,
characterized by disruptive tidal forces, high gas pressures (Bally, Stark, \& Wilson 1988; Spergel \& Blitz 1992),
and strong magnetic fields (Morris \& Yusef-Zadeh 1989), may suppress gravitational collapse in all but the most massive clouds.
Consequently, star formation  may be caused primarily by shock compression caused by cloud-cloud collisions,

The Galactic plane just outside the central molecular zone between l=1.3$^o$
and 5$^o$ contains several localized cloud complexes with unusually large velocities (Bally et al. 2010)
of $\sim$ 100-200 \kms in regions less than 0.5$^o$ ($\sim$ 75 pc) in diameter.

Therefore, we adopt for the calculation of the spectra the code SUMA
\footnote{http://wise-obs.tau.ac.il/$\sim$marcel/suma/index.htm} which simulates the physical conditions
of an emitting gaseous nebula under the coupled effect of photoionization from an external source and shocks.
Both line and continuum emission from the gas
are  calculated consistently with dust reprocessed radiation,

The shock velocities,  the preshock densities, the ionization parameters, and the relative
abundances are calculated phenomenologically.
The models are   described in Sect. 2,  the results  regarding the line spectra  
are presented in Sect. 3.
The SED of the continuum is analyzed in Sect. 4.

In particular, after
 having  unveiled the LLAGN in the GC by modelling the line spectra,
we  will  try to reproduce  consistently the SED of the  Sgr B2 observed continuum  and of  
other LLAGNs (Falcke et al 1998, Anderson et al 2004) in the radio domain.  The results are presented in Sect. 4.2.

Finally, an  interesting issue concerns the [CII]157.5/FIR  ratios, where FIR is the continuum FIR flux.
 Yasuda et al. claim, on the basis of the  low [CII]/FIR ratio in the central region of the Galaxy,
 that the dominant sources of the FIR luminosity in the GC region are not young OB stars,
but K and M giants, implying a low star-formation activity.

The [CII]  lines are emitted  from the gas, while the continuum  FIR   represents dust reprocessed 
radiation rather than  bremsstrahlung, depending on the dust-to-gas ratios.
Dust reradiation dominates  the  SED  in the FIR range.
So in this paper, we  will address the [CII]/FIR question in the GC  (Sect. 4.3) on the basis of the   
dust-to-gas ratios.

Discussion and concluding remarks follow in Sect. 5.

\section{The models}

In previous investigations of the physical conditions near the GC (e.g. Simpson 2007) the  shocks
were invoked  to explain  the line ratios and the FWHM of line profiles. 
Moreover an underlying shock generated turbulence
was confirmed  on the basis of the power spectra of the radial velocities (Paper II).

Therefore,  to model the spectra in the GC and in the disc , we will take into account  the
effects of shock wave hydrodynamics coupled to the  photoionization flux from the stars,
 or to the photoionization power-law flux from  AGN.

The  spectra observed in each position by Yasuda et al are relatively  poor in number of lines.
However, the FIR lines  presented in each spectrum, are suitable  to constrain the model physical parameters. 
Yet the calculation of the relative abundances requires some general assumptions (see Sect. 3.4).
Therefore, we cannot  select BB or  pl dominated   fluxes by the best fit of the
spectra, as was done e.g. for the Seyfert galaxy NGC 7130 (Contini et al. 2002).

In the present investigation,  we will calculate each spectrum  adopting  BB and  pl  dominated models
  (in two different sets of models),  both  reproducing as  much as possible the line ratios.
The weighted sum  of the line intensities calculated by the two models will give acceptable results 
whichever the  relative weights adopted. 

In this way, our modelling will not provide  definitive results in each position,
namely if the stars   or the active nucleus dominate the radiation field, but some interesting  results
will arise from the distribution of the parameters throughout the GC and the few disc
positions (Fig. 1).

In particular,
the emitting clouds move either towards the radiation source or outwards. Consequently,
the flux from the stars (or from the active centre)
reaches  the  shock front edge  or the edge opposite 
to the shock front, respectively.
The code SUMA  is adapted to both cases.
When the cloud recedes from the photoionizing source, the calculations
are  iterated up to convergency of results.
The geometrical thickness of the emitting nebula  plays an important role particularly 
in this case.

 We adopt a plane-parallel geometry.
The calculations start at the shock front where the gas is compressed and thermalized adiabatically,
reaching the maximum temperature in the immediate post-shock region (T $\sim$ 1.5 10$^5$ / (Vs/100 \kms)$^2$).
T decreases downstream following the cooling rate, which is calculated in each slab
downstream. This region  is cut into a maximum of 300 plane-parallel slabs 
with different geometrical widths, which are
calculated automatically, in order to  follow the temperature gradient.

The input parameters: the  shock velocity \Vs, the   preshock density \n0,
the preshock magnetic field \B0, define the hydrodynamical field. They  are  used in the calculations
of the Rankine-Hugoniot equations  at the shock front and downstream. These equations  are combined into the
compression equation which is resolved  throughout each slab in order to obtain the density profile downstream.
The  results of  line intensity calculations in each slab  depend  strongly
on the density.
A detailed description of  model calculations is given in Paper I.

The input parameters  that represent the radiation field are :  the colour  temperature of the  stars \Tef
and the ionization parameter $U$ for black body dominated fluxes,  or the power-law 
flux  from the active center $F$  in number of photons cm$^{-2}$ s$^{-1}$ eV$^{-1}$ at the Lyman limit.
if the photoionization source is an active nucleus.
Both $U$ and $F$ are calculated by radiation transfer throughout the slabs downstream.

The dust-to-gas ratio ($d/g$) and the  abundances of He, C, N, O, Ne, Mg, Si, S, A, Fe relative to H,
are also accounted for. They  affect the calculation of the cooling rate.

The dust grains are heated radiatively by  photoionization  and, collisionally, by the shock.
The distribution of the grain radii downstream
is determined by sputtering,  beginning with an initial  radius of 1.0 \mum.

The input parameters which lead to the best fit of the line ratios 
determine the physical conditions in each of the observed positions.

\section{Line spectra}

Yasuda et al. (2008) selected 28 observations from the centre and 5 observations from the disc,
which are least contaminated by the active H II regions; the observational points in the GC
region are shown in  Fig. 1.
They refer to the  region  between -3$^o$ $\leq$ l $\leq$ 4$^o$ as to the GC region; 
the disc region is outside this area.
All the observation points are close to the Galactic plane.

 Strong continuum emission and six fine-structure atomic emission lines have been detected. The lines are :
[O III] 52 $\mu $m ( $^{3}{\rm P}_{2}$  $\rightarrow$ $^{3}{\rm P}_{1}$),
[O I] 63   $\mu $m ( $^{3}{\rm P}_{1}$  $\rightarrow$ $^{3}{\rm P}_{2}$),
[O III] 88 $\mu $m ( $^{3}{\rm P}_{1}$  $\rightarrow$ $^{3}{\rm P}_{0}$),
[N II] 122 $\mu $m ( $^{3}{\rm P}_{2}$  $\rightarrow$ $^{3}{\rm P}_{1}$),
[O I] 145  $\mu $m ( $^{3}{\rm P}_{0}$  $\rightarrow$ $^{3}{\rm P}_{1}$), and
[C II] 158 $\mu $m ( $^{2}{\rm P}_{3/2}$  $\rightarrow$ $^{2}{\rm P}_{1/2}$)
in most of the GC and in a few disc regions. The intensities of the emission lines were obtained by Gaussian plus linear fitting.

\subsection{The line ratios}

\begin{figure}
\begin{center}
\includegraphics[width=0.60\textwidth]{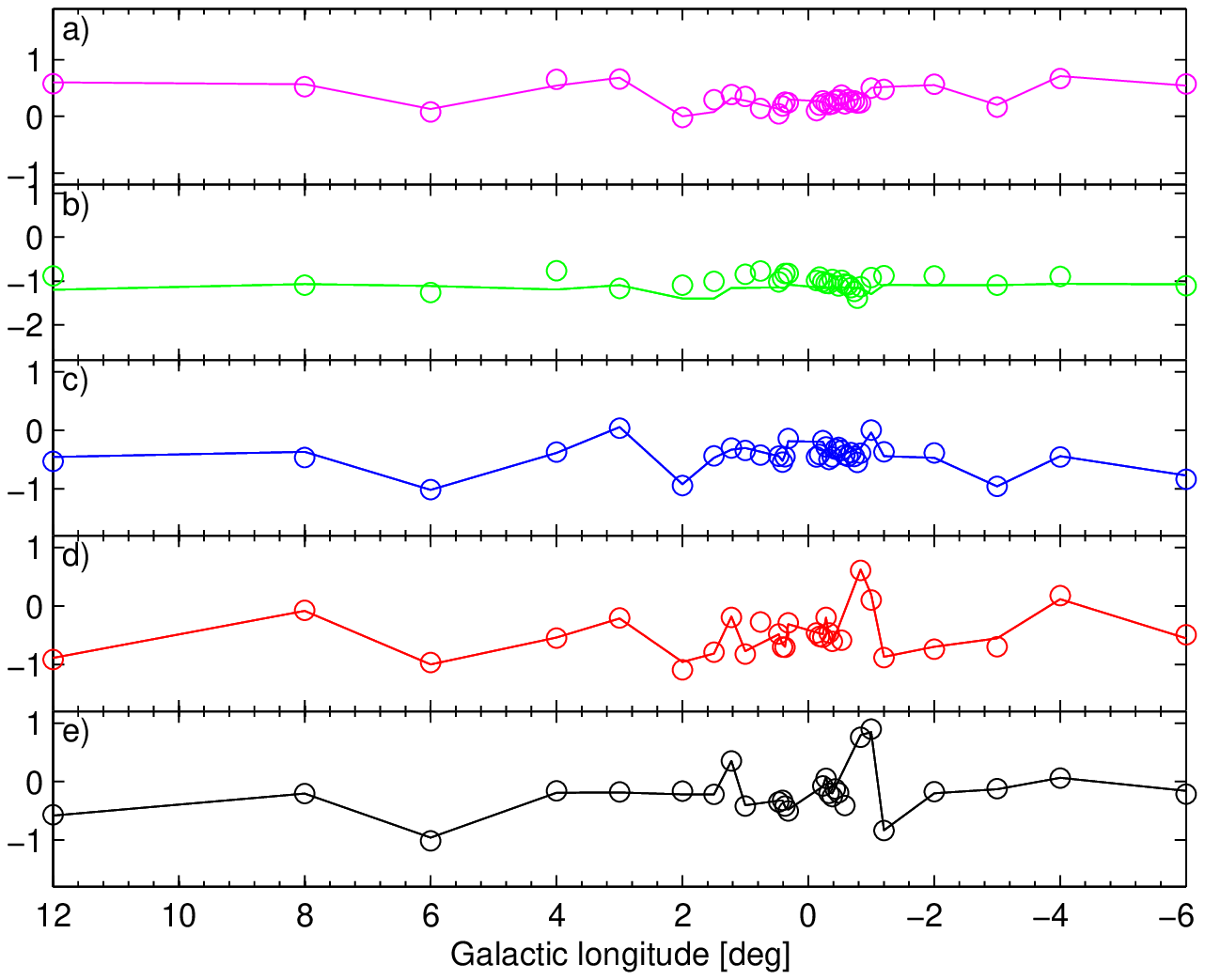}
\includegraphics[width=0.60\textwidth]{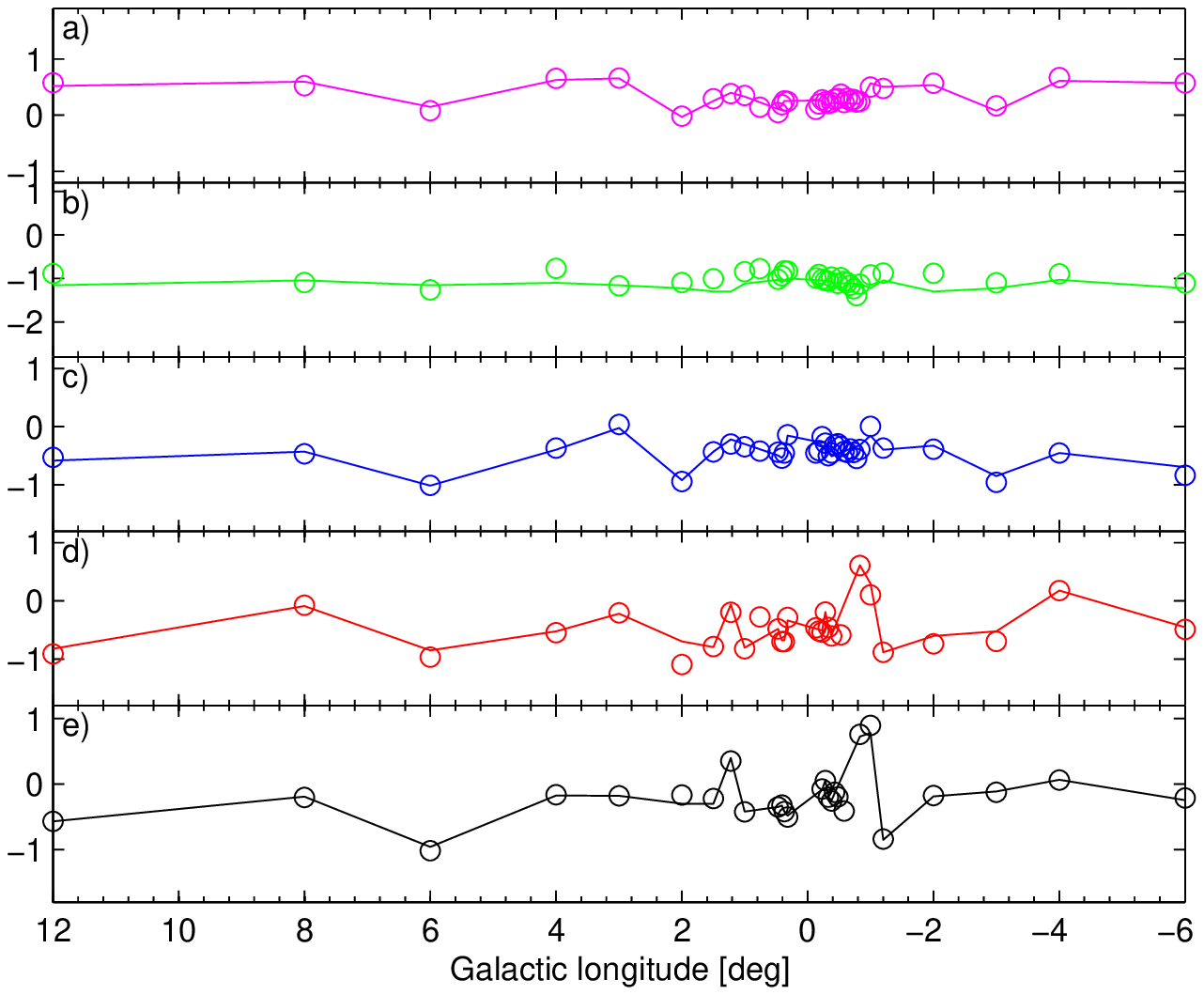}
\caption{The observed corrected line ratios (open circles) are compared with
model results (lines). Top diagrams : line ratios calculated by BB dominated
models (top) and pl dominated models (bottom). 
Black : log([OIII]51.8/[OI]63.2); red : log([OIII]88.4/[OI]63.2); blue : log([NII]121.9/[OI]63.2);
green : log([OI]145.5/[OI]63.2); magenta : log([CII]157.7/[OI]63.2).
}
\end{center}
\end{figure}

A reliable information   about the physical conditions throughout
the GC,  is obtained by modelling  line {\it ratios} rather than  {\it absolute fluxes}   
because they avoid problems of  distances, absorption etc..

In this paper we refer to  the  spectra presented by Yasuda et al (2008)
in their table 1.
We   adopt  line ratios to [OI]63 which is  a relatively strong line. 
The  observational errors of the data are rather high.
Nevertheless, we have tried to  reproduce the line ratios as precisely as possible.

Four of the six observed lines   imply oxygen ions,  therefore  the oxygen relative abundance will not
seriously affect the choice of the  physical  parameters.
We  start by   modelling  the observed line ratios, in particular  [OIII]51.8/[OI],
[OIII]88.4/[OI] and  [OIII]51.8/[OIII]88.4 by  BB and  pl dominated models
in  two different sets of calculations.

We  notice that the calculated  [OIII]51.8/[OIII]81.4 line ratios depend strongly
on the shock velocity.  Actually, \Vs  defines the postshock maximum temperature of the gas,
while the [OIII] line ratios should depend mainly on the density because they are  fine structure lines
belonging to the same triplet. (The population of the triplet levels depends mostly on the density).
The detailed modelling of the spectra leads some times to  unexpected  results.

Notice that the
 range of  \Vs, which  gives  the best fit  to the data, corresponds to  gas temperatures
downstream near the shock front  most suitable to the O$^{++}$ ion.  Throughout the  region dominated by  O$^{++}$ 
  the density trend is determined by  compression which also depends on \Vs.

The density  affects strongly the cooling rate ($\propto n^2$),  i.e.  the drop of the temperature downstream
which  determines the III/II line ratios. In Yasuda et al. spectra,  nitrogen and carbon appear  
only  through a single line, [NII] and [CII], respectively.
Therefore a reliable choice of the N/O and C/O relative abundances depends
on the   density  downstream that is constrained by  the  oxygen line ratios.

The ionizing parameter too  affects the  [OIII]51.8/[OIII]81.4 line ratios as well as the III/II line ratios.
Therefore \Vs, \n0, and $U$ ($F$ in pl models) must be calculated consistently.

Although  the observed
 [NII]/[OI] and [CII]/[OI] line ratios can be reproduced by   suitable  N/O and C/O,
the C/H, N/H and O/H relative abundances are not constrained by  Yasuda et al spectra. 
A further  constraint  is  provided  by  their  upper limits, 
namely the solar abundances (e.g. Asplund et al 2009).
In fact, the observed positions were purposly chosen to be the least contaminated by strong discrete sources.
Moreover, dust grains and molecules
which are distributed inhomogeneusly throughout  the ISM, trapping O, N, and C atoms,
may reduce the relative abundance of these species in the gaseous phase.

\subsection{Correction of the spectra}

In the first modelling  attempt, we
obtained a satisfying fit  of all the line ratios except  [OI]145.5/[OI]63.2 ($\equiv$   \roi),
both by  a power-low  flux similar to that of  AGN, and by
a black body spectrum.

The  \roi line ratios deserve a special comment.
The observed \roi   were compared with the results of  model calculations.
\roi depends inverse-proportionally on the density. Adopting the input parameters
within the ranges which give the best results for the other line ratios,
the calculations  give \roi close to 0.08  while
the observations show  $\sim$ 0.12.
The theory (Osterbrock 1988) indicates that  [OI] 145 ($^3P_0$ $\rightarrow$ $^3P_1$)/
[OI] 63 ($^3P_1$ $\rightarrow$  $^3P_2$) 
vary between 0.06 in the high density limit and 0.19  at the low density limit.
A discrepancy  by a factor of $\sim$  1.5  (and even larger in positions at -3.0, -1.0. and 2.0 deg.)  
indicates that the data should be further corrected for extinction.
The correction  affects  also the [NII] 121.9, [OI]145 and [CII] 157.7 line intensities.

We  therefore  calculated an appropriate value for the extinction A$_V$,
comparing in each position the observed  \roi
with the    value calculated at the nebula.
Adopting (Rodr\'{i}guez-Fern$\acute {\rm a}$ndez et al. 2004) : $\tau_{\lambda}$ = $\tau_{30}$(30)/$\lambda$)$^{\beta}$
(where $\tau$ is the optical depth and $\beta$ =1 for amorphous silicate grains)
A$_V$ results  $\sim$ 47 magn, higher than 30, which is generally  assumed in the GC.
However, in  positions  -3.0, -1.0, and 2.0 deg the discrepancies between
the observed \roi  (0.24, 0.8, and 0.31, respectively) and the calculated ones ($\sim$ 0.08 -  0.01)
were  still high.
Correcting  these single positions, we obtained A$_V$= 112, 266, and 174 magn , respectively, in agreement with.
Bally et al (2001) claim that  A$_V$ can be in excess of 100 magn.

All the lines observed in  each  position were corrected   adopting the calculated A$_V$ 
and the results are compared with a new cycle of calculations in Fig. 2.
We have adopted a graphical representation of the results in order to
show  the profile of the line ratios throughout the observed GC region.
The spectra  lacking one or more lines were neglected.
Fig. 2 shows that the fit of calculated to observed (corrected) line ratios is  good enough
for both BB and pl models.

Following our method (Paper I) we proceed by modelling the spectra from one
position to the next closest one adopting as an initial trial the conditions
found in the previous one. In this way we obtain  reasonable   physical conditions
and   relative abundance with minimum fluctuations.

\begin{figure}
\includegraphics[width=0.50\textwidth]{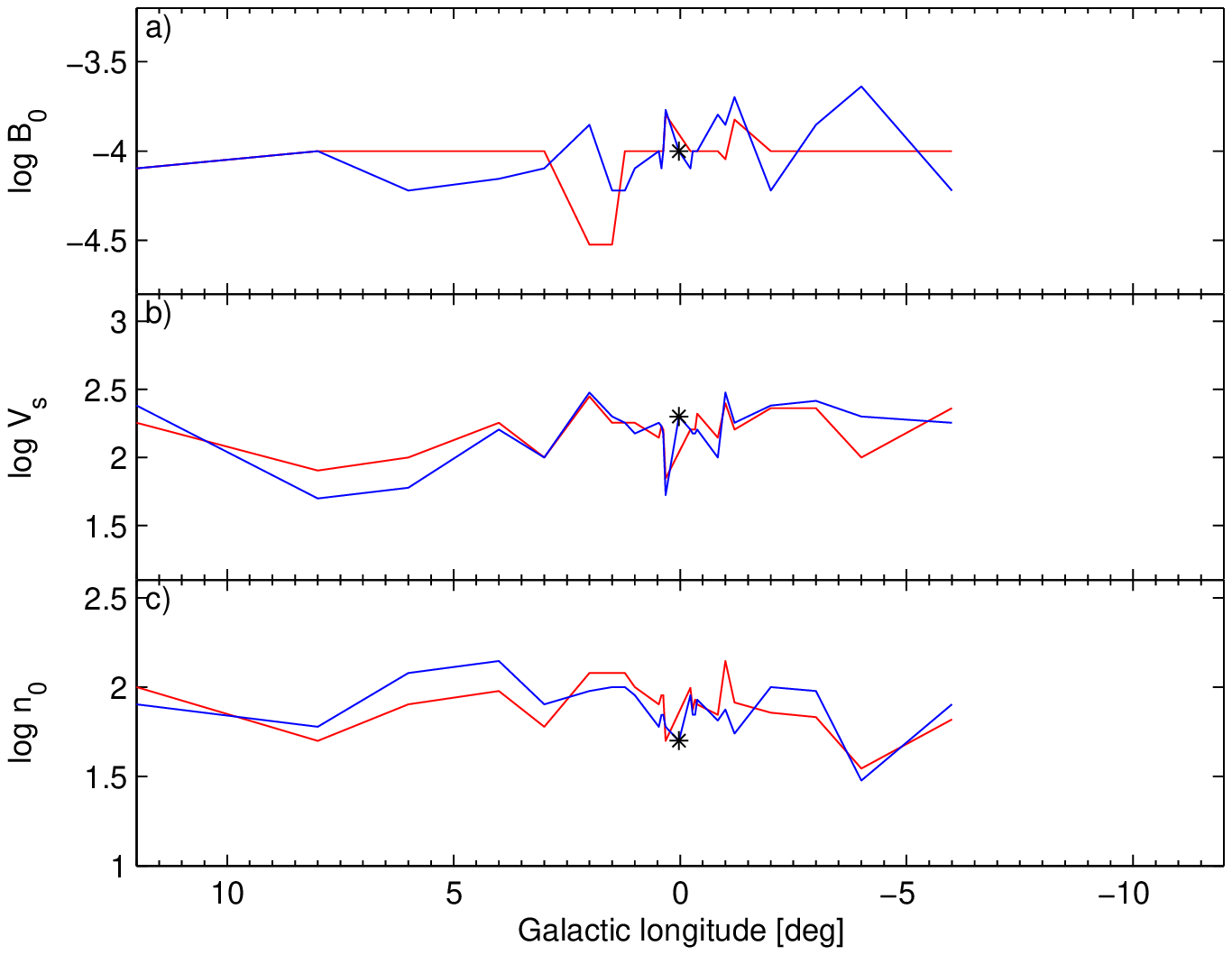}
\caption{Comparison of BB (red)  and pl (blue) models.
The input physical parameters.   
}
\includegraphics[width=0.50\textwidth]{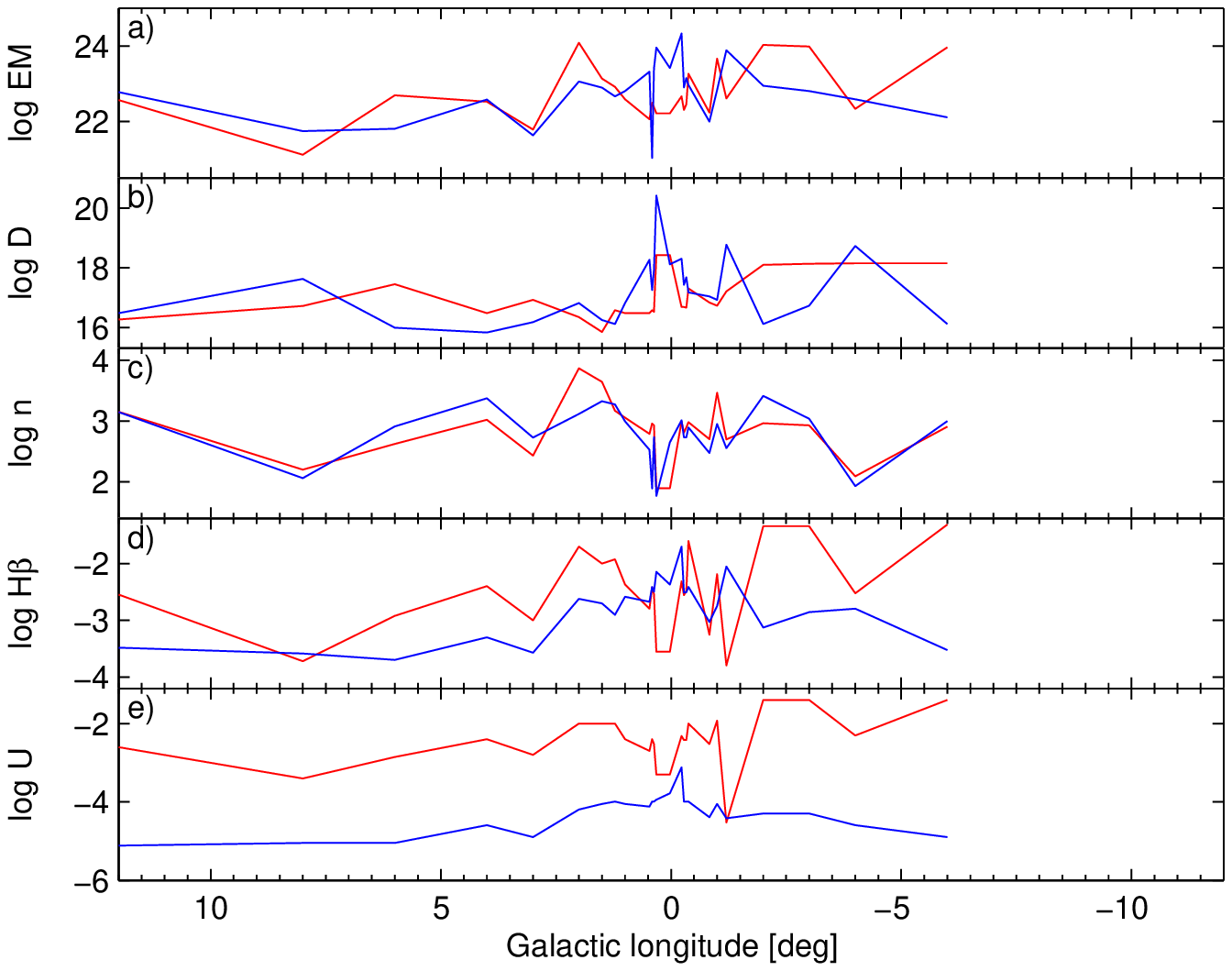}
\caption{Comparison of BB (red)  and pl (blue) models.
The emission measure (EM) in cm$^{-6}$ cm (a), the geometrical thickness of the nebula ($D$ in cm) (b), 
The downstream density (c), the \Hb flux (in \erg) (d), and the 
ionization parameters (e) (see text). 
}
\includegraphics[width=0.50\textwidth]{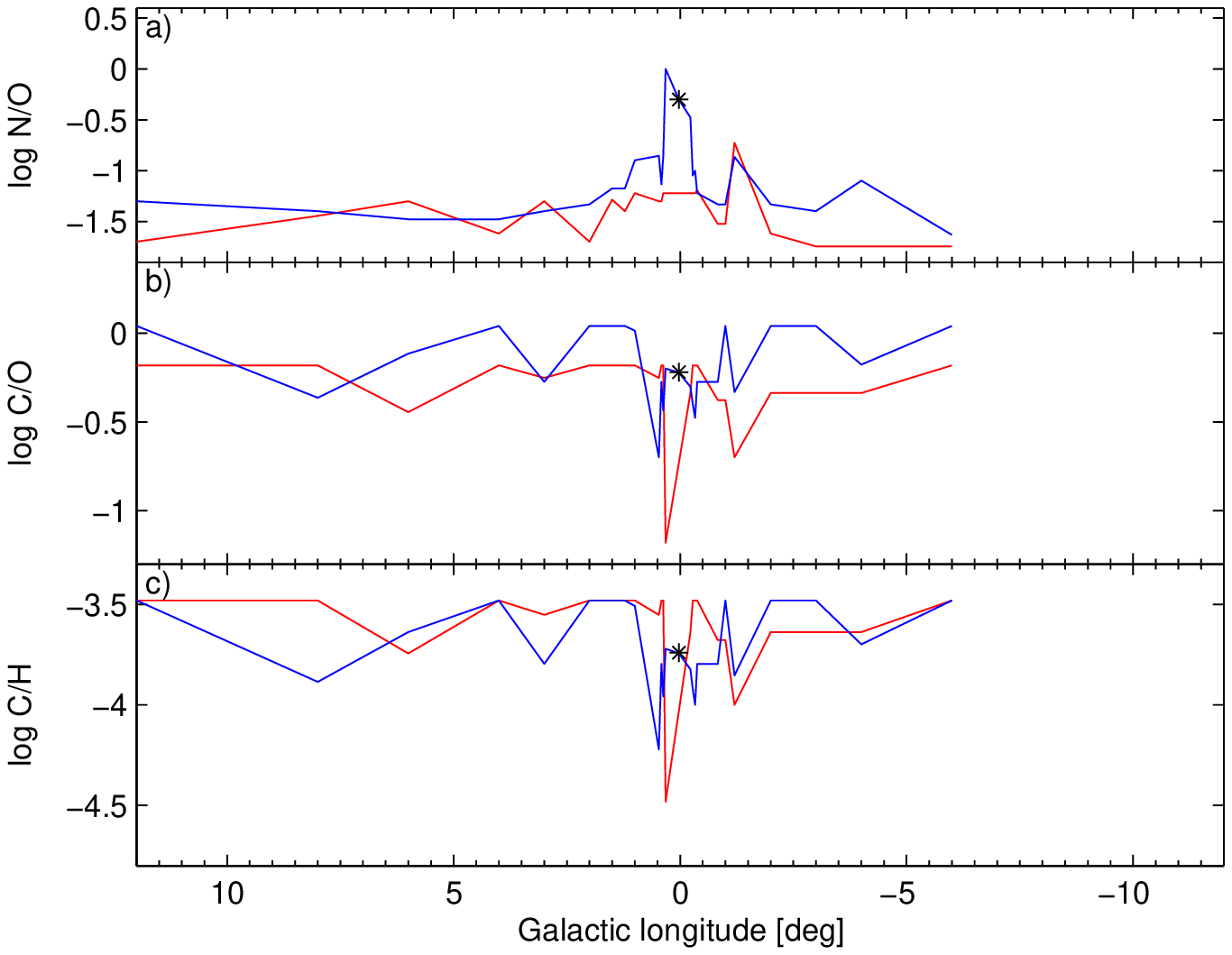}
\caption{Comparison of BB (red)  and pl (blue) models which explain the line ratios.
The relative abundances. 
}
\end{figure}

\subsection{The physical conditions throughout the GC}

The input parameters which  explain the observed line ratios
throughout the GC and the disc are presented in Fig. 3, except of $F$ which appears in Fig. 1 (bottom diagram).
In fact $F$  gives some new information  about the active centre  in the Galaxy. 
In Fig. 4 some interesting results are shown. The relative abundances appear in Fig. 5. 
The characteristic fluctuations of the parameters  in Figs. 3-5 indicate
a highly inhomogeneous medium. 

\subsubsection*{The  radiation field}

The flux $F$ adopted as the input parameter in pl models along the GC 
  is  in units of  number of photons cm$^{-2}$ s$^{-1}$ $eV^{-1}$ at the Lyman limit. 
The  spectral indices are $\alpha_{UV}$=-1.5 and $\alpha_{X}$=-0.7.

Notice that $F$ is relatively  low (7$<log(F)<10$) compared with AGN. It is
similar to  the fluxes  characteristic of  LINERs 
(Contini 1997) and LLAGNs (Contini 2004a).
In particular,  $F$ 
shows a maximum  near the Galactic centre between positions 1.0 and -1.0 deg.,  
 which clearly indicates an active nucleus.

Diagram (c) in the   bottom  panel of Fig. 4 shows  the profile of
the ionization parameters which were adopted to reproduce the spectra by BB  (red) and pl models (blue).
In the pl case  $U$  is calculated by :
$U$= ($F$/(n c ($\alpha$ -1)) (($E_H)^{-\alpha +1}$ - ($E_C)^{-\alpha +1}$)
(Contini \&  Aldrovandi 1983), where
E$_H$ is H ionization potential  and E$_C$ is the high energy cutoff,
n the density, and c the speed of light.

Comparing $U_{BB}$ ($U$ calculated by the BB dominated models) with $U_{pl}$ ($U$ calculated in the pl case)
Fig. 4 shows that $U_{BB}$ is higher than $U_{pl}$.  In fact, $U_{BB}$ 
 depends on the central cluster and  on stars throughout  the GC and in the disc. 
We suggest that the ionization
sources in the BB case  are closer to the emitting clouds than the active nucleus, 
therefore $U_{BB}$  is relatively high.
 U$_{BB}$ has not a specific trend on a large scale. 
It  shows  high fluctuations 
 between 1.00 and  -5.0 deg. 
indicating  that the flux is absorbed by  large dust complexes.
In these regions a low star formation rate was  suggested by e.g. Yasuda et al.

The ionization parameter $U_{BB}$ and the temperature of the stars  affect the spectra
in the same way. We have found that \Tef is rather constant (3.5-3.6 10$^4$ K) in agreement with
Goicoechea et al (2004), while  a  weak maximum
of 3.8 10$^4$ K  better  explain the spectra in the positions between -0.83 and -0.23 deg.

\subsubsection*{The  magnetic field}

We find a nearly constant magnetic field of 10$^{-4}$ gauss (Fig. 3).
Modelling the corrected data,  we have considered  possible
 deviations of \B0. Small scale fluctuations    can be generally
compensated by small  fluctuations of the preshock density \n0. 
In fact a high transversal \B0 prevents compression downstream.
The peak  at position -4.0 however, seems real because the preshock density  has also a minimum
and the shock velocity is low.

\subsubsection*{The densities}

The pre-shock density \n0 which appears in Fig. 3 represent the density in the medium
upstream.  Notice that BB and pl models give similar results.
The gas throughout the shock-front is compressed and thermalized.
Compression is determined by the adiabatic jump (n/\n0 =4, where n is the density)
and a further factor depends  on the magnetic field and on the shock
velocity. The calculations lead to densities downstream  which range between $\sim$ 100 and 2000 \cm3
(Fig. 4).
The electron density downstream follows the recombination of the gas,
which is accurately calculated throughout many slabs. 
   
The preshock densities which lead to the best fit of the observed line ratios
are $\leq$ 100 \cm3,  similar to those found in the regions near the GC
close to the Quintuplet Cluster and the Arches Cluster  but higher
than those found in the ISM (Paper I).
Fluctuations of \n0 in Fig. 3   show that the ISM density  between the
cloud fragments is $\sim$ 50 \cm3.
Recall, however, that the choice of the preshock density  can be affected by  the preshock
magnetic field.

The geometrical thickness of the emitting clouds ($D$) and the emission measure (EM=$\Sigma$n$^2$ $D$)
are also shown in Fig. 4.

\subsubsection*{The velocity field}

The velocities  range between a maximum of 300 \kms and a minimum of 70 \kms (BB models)
and 240  - 70 \kms for  pl models (Fig. 3).
The ranges are very similar, indicating that the shock velocity has an important role
in the interpretation of the GC spectra. 
The fluctuations  reflect  an underlying turbulent regime (Paper II).
The velocities are generally higher than those found near the GC ($\sim$ 70 \kms, Paper I)

Our modelling by BB models suggests that in the disc positions (-6.0, -4.0 deg) and close to the disc (-3.0 
and -2.0 deg),
 the clouds are receding  from the radiation source, while in the central region they move
towards the central star cluster (and the BH). 

\subsection{Relative abundances}

The [OIII]51.8 /[OI]63.2, and [OIII]88.4 /[OI]63.2 line ratios constrain the physical parameters
(\Vs, \n0, $F$ in pl models or $U$ in BB models), however, nitrogen and carbon appear only
 throughout one line, [NII]121.9 and [CII]157.7, respectively.
Our modelling determines the N/O and C/O abundance ratio in each observed position (Fig. 5),
but to obtain the abundances relative to H, at least one  of the  C/H, N/H, and O/H, should 
be  constrained.

The observations by Yasuda et al cover typical regions in the GC least
contaminated by strong discrete sources, namely the ISM  along the GC extended region.
Here gas and dust are strongly coupled.
We consider for instance, that C can be locked in diatomic molecules  like CO, CN, CS 
 and in carbon grains, generally in PAHs and other species (see Amo-Baladr\'{o}n et al 2011).
PAH are very small grains which are easily sputtered throughout shock fronts
even for shock velocities $<$ 150 \kms. 
C/H (Fig. 5) is close to  solar in agreement with  destruction of grains by
 harsh radiation
from the star clusters and from the AC as well as by sputtering.
So we set the  C/H upper limit at the solar value (Fig. 5, bottom diagram).
 C/H variations show in particular the distribution of dust.

A rapid comparison between Fig. 3 and Fig. 5  shows that the  
fluctuations in the profiles of the
physical parameters are also present in the  C/H, C/O, and N/O relative abundances.
The C/O abundance ratio  calculated by BB models is   lower than solar ($\sim$ 0.5)  
in  positions near the very centre and nearly solar in the other positions.
 C/O calculated by pl models is  higher  ($\leq$ 1.58)  than solar throughout  most of the GC.
Theoretically, this depends on the exponential and power-law of the radiation flux, respectively.
Physically, this discrepancy shows that different conditions coexist in each of the observed position.

The calculated N/O relative abundance ($\sim$ 0.08 and $\sim$ 0.03 in average in the pl 
and BB models, respectively) is lower that solar ($\sim$ 0.25)
throughout most of the GC region, except  in the
very center.  Here N/O reaches a value of  $\sim$ 1, indicating that oxygen  is strongly depleted from 
the gaseous phase. Unfortunately,   Yasuda et al spectra do not contain  Si lines,
which could  reveal whether dust in the GC is composed by silicate grains.
C/H is rather low in the same positions indicating that CO molecules 
are present. The velocities are $\leq$ 100 \kms,
then most of C and O  can be also trapped into dust grains, while  molecules such as NH$_3$, NH,
etc, are destroyed by  radiation from the black hole.

Metallicity enhancement of CNO could derive from classical nova explosions.
However, they are easily washed  out by dust adsorption in the surroundings.

In order to better constrain the N/O relative abundance we select one of the spectra presented by
Goicoechea et al (2004, table1) which responds to two criteria :
1) it contains data  of all the observed lines including [NII] and [NIII] and 
2) it is placed in the GC region (l=0.025 deg, $\sim$ 3.5  pc)) 
at a latitude b=0 deg. 
We have corrected the spectrum for A$_V$=250 mag as indicated by Goicoechea et al (2004).
The comparison between observed (corrected) and calculated line ratios to [OI]63.2 appears in Table 1.
The model is calculated adopting \n0=50 \cm3, \Vs=200 \kms, \B0=10$^{-4}$ gauss, 
$F$=1.3 10$^9$ photons cm$^{-2} s^{-1} eV^{-1}$
at the Lyman limit, C/H=1.8 10$^{-4}$, C/O=0.6, and N/O=0.5. These values are inserted in 
Figs. 3-5 as  black asterisks. The low C/H relative to the solar value (2.7 10$^{-4}$, Asplund et al 2009)  
is in agreement with  
the strong  extinction in the  very centre of the Galaxy,   and with the high dust-to-gas ratios  (see Sect. 4).
 
\begin{table}
\caption{comparison with Goicoechea et al (2004) spectrum}
\begin{tabular}{ l l l l l } \\ \hline \hline
\ line ratios to [OI]63.2   & obs. & calc. \\ \hline
\ [OIII]51.8                &1.2  & 1.1 \\
\ [NIII]57.3                &0.81 & 0.80 \\
\ [OIII]88.36               &0.5 & 0.65 \\
\ [NII]122.                 &0.12 & 0.18\\
\ [OI]145.5                 &0.08 & 0.09\\
\ [CII]157.7                &0.38 &0.4  \\ \hline
\end{tabular}
\end{table}

\section{The SED of the continuum}

The continuum spectra presented by Yasuda et al (2008, fig. 2) cover the  wavelength range $\sim$ 50-200 \mum.
The models which  explain the line spectra in each position
are adopted to calculate consistently  the free-free and free-bound emission
flux from the gas and  IR reprocessed radiation  from dust grains.
The continuum from the gas and reradiation from dust appear in Figs. 6 and 7 as two distinct curves
for each model.

\subsection{Temperature of the dust grains}

\begin{figure}
\includegraphics[width=0.50\textwidth]{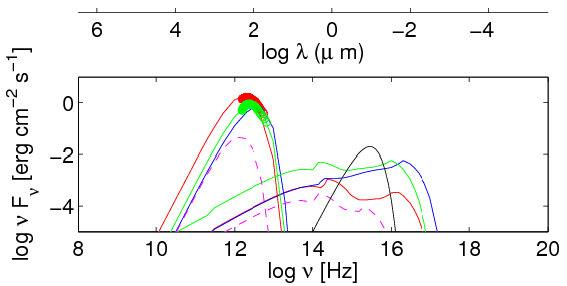}
\caption{Comparison of calculated continuum SEDs with LWS data (Yasuda et al 2008).
Red  circles: the data from Yasuda et al. in the Galactic central region
Green  circles: the data from Yasuda et al. in the Galactic disk.
   Red   solid lines : pl model  at position l=0.025 deg.
Blue solid lines : pl model  at position l=-3. deg.
 Green   lines :BB  model  at position l=12. deg.
Magenta lines : BB model at position l=8. deg.
Black line : the black body corresponding to T$_{eff}$=35000 K.
}

\includegraphics[width=0.50\textwidth]{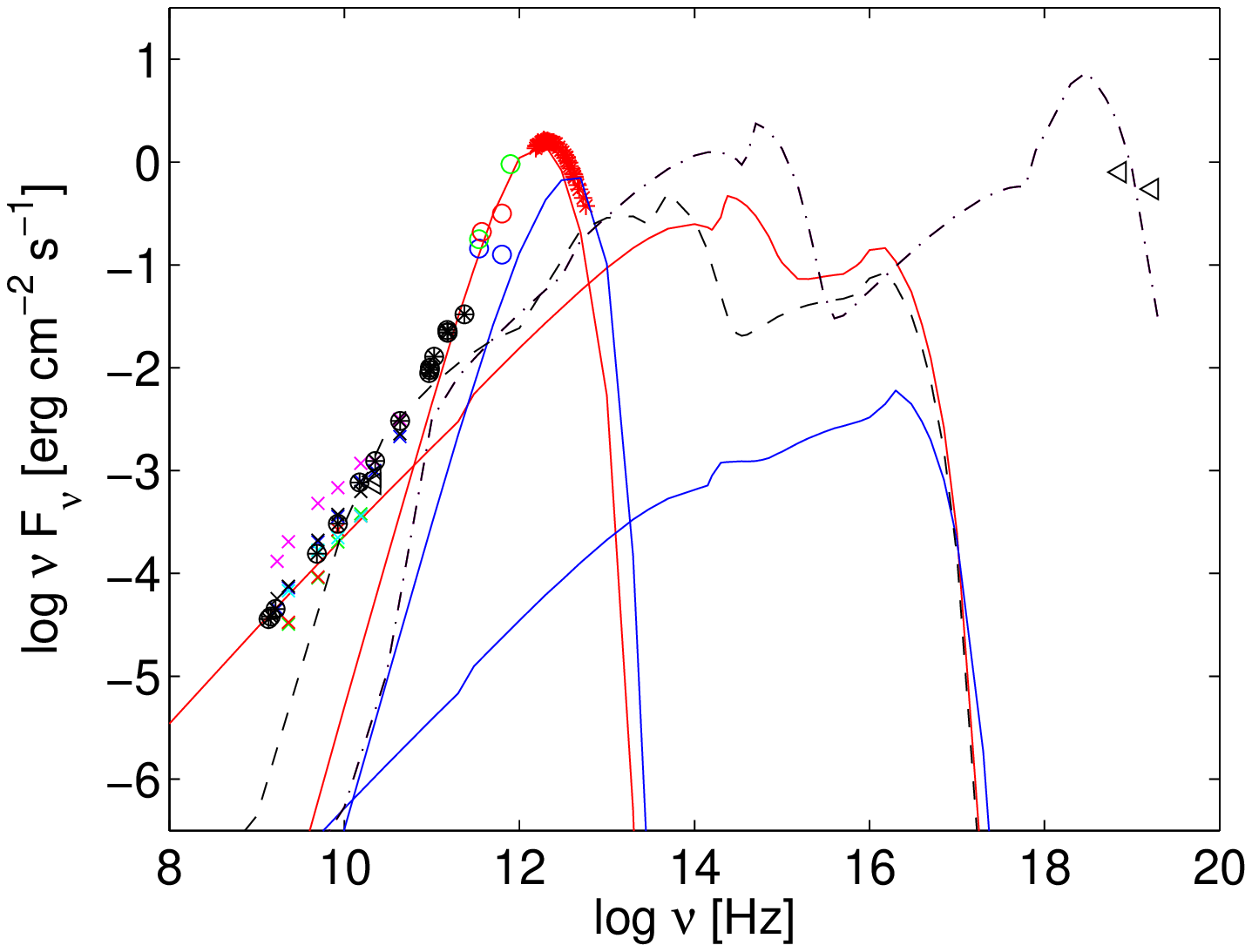}
\caption{Comparison of calculated continuum SEDs with data.
   Red   solid lines : pl model  for position l=0.025 deg.;
black dashed lines : pl model calculated by \n0= 3 10$^4$ \cm3 and \Vs=200 \kms;
black dot-dashed lines : pl model calculated by \n0= 6 10$^5$ \cm3 and \Vs=3000\kms;
blue solid lines : pl model at position -3 deg..
Anderson et al (2004) data : black X : NGC 3147; cyan X : NGC 4579;  blue X : NGC 4203;
red X : NGC 4168; green X : NGC 4235; magenta X : NGC 4450.
Pierce-Price et al. (2000) data in the radio : open circles.
Falcke et al (1998) data : filled circles.
Yasuda et al data in the IR : red asterisks.
Bird et al. (2007) data for Sgr A* in the X-range : black  open triangles.
Petrov et al (2007) data for Sgr A* in the radio range : black open triangles
}
\end{figure}

\begin{figure}
\includegraphics[width=0.50\textwidth]{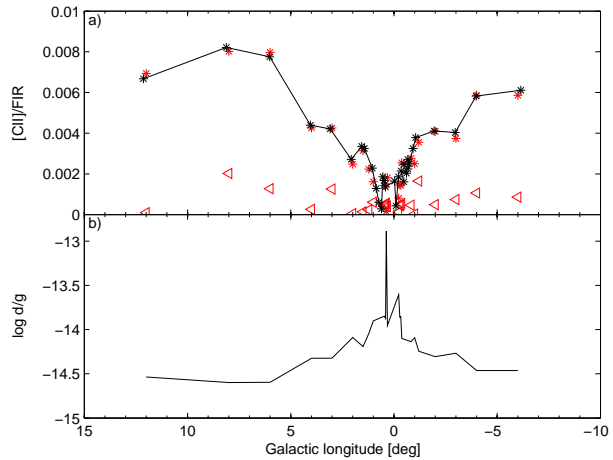}
\caption{Top panel : comparison of  model calculations (red  triangles) with [CII]/FIR data (black asterisks)
adapted from Yasuda et al.(2008, fig. 4)
adopting [CII] and FIR values summed up from BB and pl models. Red asterisks represent the adjusted results (see text).
 The $d/g$ ratios appear in the bottom panel.
}
\end{figure}

The  grains are heated by radiation and by mutual collisions
with the gas and their temperature is calculated in each of the slabs downstream at each 
of the observed positions.
The temperature of the grains depends also on the  grain radius, therefore on sputtering.
Integration throughout the slabs  for each model leads to the dust  reprocessed
radiation curve  which covers a range of frequencies larger than a simple Planck function.

In Fig. 6  we  compare  model calculations with the LWS data
of  the continuum SED in   two  characteristic positions 
presented  by Yasuda et al (2008, fig. 2), one in the GC and the other in the disc. 

We have selected two  models  from the modelling of the line spectra in the GC, one at positions 0.025 deg. 
(from Goicoechea et al. 2004)  
close to the very centre, and the other  at position l=-3. deg, at the 
  centre-disc limit, to model the observed data (red circles)  in the GC. 
Both models are calculated adopting a pl flux from the active centre.
The calculated maximum grain temperature is \Tgr=41 K for the model closest to the GC  and a maximum of 52 K
for dust at position l=-3 deg. The grain maximum temperature is proportional
to the shock velocity. 
  A grain  radius \agr =1. \mum is adopted.

The  green  circles refer to the observations from the  disc.
The model fitting the line spectrum at position l=12 deg reproduces the Yasuda et al data satisfactorily.
The model is calculated adopting   a BB photoionizing  
 radiation flux from  the  stars with \Tef=35000K.
The temperature of the silicate grains  calculated in the slabs downstream
covers the  20 - 44 K temperature range.

The  calculated dust temperatures are in agreement with Yasuda et al results (14-20 and 25-40 K from source to source).
However, the relatively small frequency range covered by the data does not  constrain the relative intensity  
between the dust reradiation peak and  the bremsstrahlung (see Paper I).
We adopted in these models a dust-to-gas ratio of 1-5 10$^{-14}$ by number corresponding 
to $\sim$ 4 10$^{-4}$ - 2 10$^{-3}$ by mass for silicates.

Fig. 6 shows that the two  observed curves in the FIR (Yasuda et al, fig. 2, top and bottom diagrams)
 can be fitted by   the continuum spectra
emitted from different positions. The shock velocity is the critical parameter. 

We have added in Fig. 6 the black body radiation corresponding to  \Tef = 35000 K
from the stars by a free Y-axis scale.

\subsection{The radio-FIR range}

We present in Fig. 7 the modelling of the continuum SED on a larger frequency range.
We have added to the continuum SED observed in the central region (Yasuda et al 2008, fig.2, top diagram) 
the following  data :

1)  observed by SCUBA in the mm range
by Pierce-Price et al (2000) (red open circles), observed by Serabyn et al (1997) with the
Caltech submillimeter Observatory and JCMT (green open circles) , and observed by Zylka et al (1995)
(black open circles). 
 Pierce-Price et al (2000) who investigated variability,  presented  data in the  range between 700 and 2000 \mum.
The data are roughly  reproduced by  dust FIR reradiation.

2)  observations of Sgr A* corresponding to  centimeter to
millimeter wavelengths  obtained by   the VLA, the Berkeley-Illinois-Maryland Array (BIMA), 
the Nobayama 45 m. and the IRAM 30 m. telescope. 
The data were presented  by Falcke et al. (1998), who  explained the millimeter excess by
the presence of an ultracompact component of relativistic plasma with the size of a few
Schwarzschild radii near the black hole. 
Indeed the data cannot be reproduced by the slope of  the dust reradiation curve, nor by the
bremsstrahlung corresponding to the densities suitable to the observed line ratios,  but they show that
self-absorption of free-free radiation should be accounted for.

3) For comparison we show in  Fig. 7 the SEDs of LLAGNs (NGC 3147, NGC 4579, NGC 4203,
NGC 4168, NGC 4235, and NGC 4450) in the radio range presented by Anderson et al (2004, fig. 3).

During the past decade
advection-dominated accretion flows  have been used to explain the 
SED of Sgr A* and other LLAGNs 
(Narayan \& Yi 1995; Narayan, Yi \& Mahadevan 1995).
Namely, a hot plasma develops in which the protons decouple from the electrons,
allowing the protons to carry most of the energy released by the accretion process
into the black hole, limiting the amount that is radiated away.
Hot thermal electrons generate synchrotron radiation that is self-absorbed, resulting
in a radio spectral index of $\sim$ 0.4 (Mahadevan 1997).
Bremsstrahlung and inverse Compton processes generate additional emission, producing a
relatively well-defined spectrum from radio through X-ray frequencies
(Anderson 2004).

Following our model for the GC (Papers I and II) the clouds are fragmented by the
underlying turbulent regime, so many different physical conditions may coexist  in different
positions, in particular, close to Sgr A*.
For instance,  the detailed modelling  of  the LLAGN NGC 4579 (Contini 2004a) confirms 
that self-absorption of free-free radiation is crucial
to reproduce the trend of the radio data.

Self-absorption  requires a cloud component with densities higher
than those found to reproduce the FIR spectra  close to Sgr A*. 
The model  which fits the  Sgr A* radio continuum SED  between 10$^{10}$ and 10$^{11}$ Hz 
is calculated by  \Vs=200 \kms , \n0=3 10$^4$ \cm3 and log $F$=9.1.
At  lower frequencies the observations recover the trend of the bremsstrahlung 
emitted downstream of clouds with  lower densities. At higher frequencies the data
follow the dust FIR reradiation trend.

Other clouds with shock velocities
 up to 3000 \kms explain the X-ray data. They coexist in the very centre of the Galaxy, but are
 closer to the active nucleus  than the low velocity gas.

\subsection{Dust-to-gas ratios}

We investigate the [CII]157.7/FIR  ratios presented by Yasuda et al. (2008, fig. 4).  
First we  check  whether in  regions close to the GC dust reprocessed radiation or bremsstrahlung
dominate  the FIR continuum.
In Paper I we calculated the physical  conditions of gas and dust in the  regions photoionized by the
stars   near the GC
and collisionally heated by  shock fronts.
Gas and dust are coupled collisionally and by the mutual   heating and cooling   of gas  and dust. The grains
are affected by sputtering, evaporation, etc.
Shock velocities, which  directly determine the frequency range of   dust reprocessed  radiation
(Contini  et al. 2004),
were  constrained by the  FWHM of line profiles in the observed spectra.
By modelling the SED of the continuum it was found that indeed dust reradiation dominates the FIR.

We will refer  to [CII]/FIR, neglecting the [OI]63.2/FIR ratio, because [CII]/[OI] is rather constant 
(Fig. 2, dagrams a)) and  almost always $>$1.
Anyway, Fig. 6
shows that both the  FIR lines  appear in the range  dominated by  dust reradiation.
We present Yasuda et al ISO data (black asterisks)  in Fig. 8 (top panel)  as well as  model  results 
(red open triangles).

 The [CII] lines and the integrated
FIR continua were obtained by summing up the results of BB and pl models, adopting the same relative weights
for the two models. The  calculations were repeated in all the observed positions.
Recall that the FIR continuum is highly dominated by dust reprocessed radiation,
while  the [CII] lines are emitted by the gas.
The temperature of the grains  and the line intensities emitted by the gas are calculated consistently in each slab downstream.

The dust-to-gas ratio determines directly the relative intensity   of the  dust reradiation peak  to the bremsstrahlung.
Changing the $d/g$ parameter by a factor $\leq$  100  in  a model  would affect
the   line ratios by very small factors ($<$1\%), while large  $d/g$  alterations  can lead to a  drastic 
 different situation in shock dominated regimes, e.g. from a  non radiative to a radiative shock (Contini 2004b).

Fig. 8 shows that, even if the trend of the models  reproduces qualitatively  the data,  the fit
is not  satisfying.
Recall that the  FIR fluxes are $\propto$  $d/g$  and that
 the models were calculated adopting $d/g$= 10$^{-14}$ by number ($\sim$ 4. 10$^{-4}$ by mass). 
Therefore  to increase the [CII]/FIR ratios, the calculated FIR  fluxes should be divided by 
an adjusting factor f$_{d/g}$  in each position.  
The [CII]/FIR  calculated ratios best fitting the data (red asterisks) were  re-calculated adopting
  $d/g$= 10$^{-14}/f_{d/g}$.
The results are presented  graphycally  in  the bottom panel of Fig. 8.
The profile of $d/g$ throughout the GC shows that there is an accumulation of
dust in the very centre.

Yasuda et al (2008, fig. 4)   discuss the [CII] /FIR  minimum between 
-1 and 0 deg. on the basis of the FUV radiation, a spectrally soft radiation flux, and high gas densities.
 All these parameters are accounted for by our models and lead to consistent results.

\section{Discussion and concluding remarks}

Goicoechea et al (2004) claim that  the coupled effect of collisional and  radiative-type heating mechanisms 
{\it  seem the rule in Sgr B2 and
in the bulk of GC molecular clouds} observed by ISO (Rodr\'{i}guez-Fernandez et al 2004). 
Cloud complexes
near l=1.3, 3 and 5 deg  have velocities of 100-200 \kms in regions less than 0.5 deg ($\sim$ 75 pc) 
in diameter (Dame et al. 2001); Listz 2006, 2008). Bally et al. (2010) notice that   the cloud complexes may 
trace the locations where gas is entering the
dust lanes at the leading edges of the bar in the center of  the Galaxy  passing through a series of shocks (Listz 2006), 
or dust lanes along the bar's leading edge  which is seen nearly end-on.

A complex network of filaments created by fragmentation   at the shock fronts was  found near the GC
by  modelling the near-IR spectra in Papers I and II.

However, low velocity shocks (H\"{u}ttemeister et al 1995, Mart\'{i}n-Pintado et al 1997)
are not the only heating mechanism, because a  
 dense stellar population  in the central cluster and throughout the GC heats and ionizes gas and dust.

The modelling of the FIR spectra observed by Yasuda et al (2008) throughout the very centre of the Galaxy and in some disc positions,
 shows that the physical conditions and the relative abundances  fluctuate
within reasonable limits.
For instance, the velocities range between 50 and 300 \kms, the gas preshock densities within 30 and 100 \cm3,
and the ionization parameters (referring to the BB flux from the stars)  range between  2 10$^{-4}$ and 2 10$^{-2}$.

The calculation of the FIR spectra adopting a power-law  flux, characteristic of AGN, leads to interesting results.
First, the LLAGN character of the Galaxy is confirmed by  a  pl  maximum  radiation  of 
$F$ $\sim$ 6 10$^9$  photons cm$^{-2}$ s$^{-1}$ eV$^{-1}$  close to the Sgr A* position (Fig. 1, bottom diagram).
The maximum $F$ flux is very similar to that found by consistent modelling of  NGC 4579 ($\leq$ 10$^{10}$ Contini 2004a) 
and at least by a factor of 100 lower than in AGNs.

Fig. 4  shows that the corresponding ionization  parameter $U_{pl}$  
 decreases with distance from the centre, while the ionization parameter  ($U_{BB}$) calculated by BB dominated 
models has no specific trend, but a fluctuating profile.
 $U_{BB}$ is due to
radiation from the stars,  adopting that a temperature  \Tef $\sim$ 36000 K 
permeates the inhomogeneous medium in the GC.  
Moreover,  $U_{pl}$,  is lower by a factor of $\sim$ 100 than $U_{BB}$.
This result suggests that in the observed positions, 
 the stars  should be located closer to the emitting gaseous clouds than the active nucleus.   
However, Yasuda et al observations refer to regions {\it least contaminated by strong discrete sources}.
This strengenths the AGN  presence in the GC.

Even if the AGN in the Galaxy is weak and the lines in the UV-optical range are invisible,
we have calculated the \Hb absolute flux throughout the GC by both BB and pl models.  Fig. 4 shows that, in a small region
between  about -0.2 and 0.2 deg,
the   \Hb  flux  would trace the LLAGN rather than the central star cluster.

Reprocessed radiation from the dust which is coupled to the gas throughout the shock front  peaks at $\sim$ 100 \mum. 
The  FIR SED of the continuum flux from
 the very centre is different from that emitted from the positions near the disc,  because it depends  on the physical 
conditions of the gas, in particular on the shock velocity.

The dust-to-gas ratios calculated in the GC  are 10$^{-15}$ $\leq$ $d/g$ $\leq$ 10$^{-13}$, never reaching, however,
the values characteristic of IR luminous galaxies (Contini \& Contini 2007) even close to the centre. 
The $d/g$  were derived  from the fit of the observed [CII]/FIR ratios.

The modelling of the continuum  near the Sgr A* position  shows that the  radio SED results from the
contribution of  clouds with different densities and velocities.  A   pre-shock 
 density 
 \n0 $\sim$ 3 10$^4$ \cm3  combined with   \Vs=200 \kms   leads to densities downstream   $\sim$ 5 10$^6$ \cm3, high enough 
for  self-absorption  of free-free radiation in the radio range. 
Such  high densities were observed in the circumnuclear disc (Amo-Baladr\'{o}n et al.  2011).
At  low frequencies ($ <$ 10$^{10}$ Hz) the observations recover the  trend of the bremsstrahlung emitted downstream of clouds 
with  lower densities.

 The radio flux is variable on scales of weeks to months.
 While syncrotron self-Compton is an important component, thermal bremsstrahlung
from the accretion flow plays an important role (Falcke, H. \& Markoff, S. 2000).
The variability in the EUV to X-ray band is correlated to variability in the sub-mm. The scale of weeks and months could
be explained by the  high fragmentation of the medium crossed by the  shock front.

High velocity  components are revealed by the X-ray data (Bird et al 2007).
Observations from the Sgr A* source are explained by high-velocity shocks (\Vs $\sim$ 3000 \kms) 
characteristic of jets.
 This velocity corresponds to a postshock temperature of $\sim$ 10$^8$K. The observation of the 6.7 keV 
helium like Fe line (Sunyaev et al. 1993) argues for thermal bremsstrahlung as a possible emission mechanism, However,
the age of the source cannot be more than  $\sim$ 3 10$^4$ years (Chevalier 1982, adopting a radius
of $\sim$ 400 pc) if there is not a continuous heating mechanism.

The high velocity could  be the  shock velocity accompanying  the blast wave from a supernova outburst,  
In fact the X-ray flux is variable and decaying.
However,
the hypothesis of a supernova origin in the GC is excluded by Koyama et al (1996)
because there is no indication of repeated supernova outbursts.



\section*{References}

\def\ref{\par\noindent\hangindent 20pt}

\ref Aitken, D.K. et al. 2000, Apl, 534, L173
\ref Amo-Baladr\'{on}, M.A., Mart\'{i}n-Pintado, J., Mart\'{i}n, S. 2011, A\&A, 526, A54
\ref Anderson, J.M., Ulvestad, J.S., Ho, L.C. 2004, ApJ, 603, 42
\ref Asplund, M., Grevesse, N., Sauval, A.J., Scott, P. 2009, ARA\&A, 47, 481
\ref Bally, J.; Stark, A. A.; Wilson, R. W.; Henkel, C.	 1988, ApJ, 324, 223	
\ref   Bally, J. et al. 2010, ApJ, 721, 137
\ref  Bird, A.J. et al. 2007, ApJS, 170, 175
\ref Chevalier, R.A. 1982, ApJ, 259, L85
\ref  Contini, M. 1997, A\&A, 323, 71	
\ref  Contini, M.  2004a, MNRAS, 354,  675
\ref  Contini, M.  2004b, A\&A, 422, 591
\ref Contini, M. 2009 MNRAS, 399, 1175, Paper I
\ref Contini, M., Aldrovandi, S.M.V. 1983, A\&A, 127, 15
\ref Contini, M., Contini, T.  2007, AN, 328, 953
\ref Contini, M., Goldman, I. 2010, MNRAS.tmp.1752, Paper II
\ref Contini, M., Viegas, S.M., Campos, P.E. 2003, MNRAS, 346, 37
\ref Contini, M., Viegas, S. M., Prieto, M. A. 2002 A\&A,  386, 399
\ref Contini, M., Viegas, S. M., Prieto, M. A. 2004, MNRAS, 348, 1065 
\ref Contini, M., Radovich, M., Rafanelli, P.,  Richter, G.M. 2002, ApJ, 572, 124
\ref Cotera, A.S.; Erickson, E.F.; Colgan, S.W.J.; Simpson, J.P.; Allen, D.A.; Burton, M.G. 
 1996, ApJ, 461, 750	
\ref Cotera, A. S., Colgan, S. W. J., Rubin, R. H., Simpson, J. P. 2005,
ApJ, 622, 333
\ref Dame, T. M.; Hartmann, D.; Thaddeus, P.	 2001, ApJ, 547, 792	
\ref Eisenhauer, G. et al. 2005, ApJ, 628, 246
\ref Falcke, H., Goss, W.M., Matsuo, H., Teuben, P., Zhao, J-H., Zylka, R. 1998, ApJ, 499, 731 
\ref Falcke, H.  Markoff, S. 2000, A\&A, 362, 113
\ref Genzel, R., Eisenhauer, F., Gillessen, S. 2010, Rev. Mod. Phys., 82, 3121
\ref Ghez, A.M. et al. 2003, ApJ, 586, L127
\ref Ghez, A.M. et al. 2005, ApJ, 620, 744
\ref Goicoechea, J.R., Rodriguez-Fernandez, N.J., Cernicharo, J. 2004, ApJ, 600, 214
\ref H\"{u}ttemeister, S., Wilson, T. L., Mauersberger, R., Lemme, C., Dahmen, G.,  Henkel, C. 1995, 
A\&A, 294, 667
\ref Koyama, K.; Maeda, Y.; Sonobe, T.; Takeshima, T.; Tanaka, Y.; Yamauchi, S., 1996, PASJ, 48, 249
\ref Lis, D. C.; Serabyn, E.; Zylka, R.; Li, Y.	 2001, ApJ, 550, 761	
\ref Liszt, H. S. 2006, A\&A, 447, 533
\ref Liszt, H. S. 2008, A\&A, 486, 467	
\ref Mahadevan, R. 1997, ApJ, 477, 585
\ref Mart\'{i}n-Pintado, J., de Vicente, P., Fuente, A.,  Planesas, P. 1997, ApJ, 482, L45 
\ref McKee, C.F., Tan, J.C. 2003, ApJ, 585, 850
\ref Morris, M.; Yusef-Zadeh, F.  1989, ApJ, 343, 703	
\ref Nagata, T., Woodward, C.E., Shure, M., Kobayashi, N. 1995, AJ, 109, 1676
\ref Narayan, R., Yi, I. 1995, ApJ 444, 231
\ref Narayan, R., Yi, I., Mahadevan, R. 1995, Natur, 374, 623
\ref Osterbrock, D.E. 1988 in 'Astrophysics of gaseous nebulae and active galactic nuclei',
University Science Books
\ref Petrov, L., Hirota, T., Honma, M., Shibata, K. M., Jike, T., Kobayashi, H.  2007, AJ, 133, 2487
\ref Pierce-Price, J. S.  et al.  2000, ApJ, 545, L121
\ref Rodr\'{i}guez-Fern\'{a}ndez, N. J.; Mart\'{i}n-Pintado, J.; Fuente, A.; Wilson, T. L. 2004, A\&A, 427, 217	
\ref Sch\"{o}del, R., Bower, G,C., Muno, M.P., Nayakshin, S., Ott, T.
2006, Journal of Physics: Conference Series, Volume 54, Proceedings of "The Universe Under the Microscope - Astrophysics at High Angular Resolution", held 21-25 April 2008, in Bad Honnef, Germany. Editors: Rainer Schoedel, Andreas Eckart, Susanne Pfalzner and Eduardo Ros, pp. (2006).
\ref Sch\"{o}del, R. et al. 2003, AN, 324, 535 
\ref Serabyn, E., Carlstrom, J., Lay, O., Lis, D. C., Hunter, T. R.,  Lacy, J. H. 1997, ApJ, 490, L77  
\ref Simpson, J P.; Colgan, S. W. J., Cotera, A. S., Erickson, E. F.,
Hollenbach, D. J., Kaufman, M. J., Rubin, R. H. 2007, ApJ, 670, 1115
\ref Sunyaev, R.A., Markewitch, M., Pavlinsky, M. 1993, ApJ, 407, 606
\ref Spergel, D. N.; Blitz, L.	 1992 Natur., 357, 665	
\ref Yasuda, A., Nakagawa, T., Spaans, M., Okada, Y., Kaneda, H. 2008, A\&A, 480, 157
\ref Zylka, R., Mezger, P. G., Ward-Thompson, D., Duschl, W. J.,  Lesch, H. 1995, A\&A, 297, 83


\end{document}